\documentclass[aps,pre,reprint,groupedaddress]{revtex4-2}

\usepackage{amsfonts,amssymb,booktabs,makecell,graphicx,subfigure,multirow,enumerate}
\begin{document}

\title{Error Processing of Sparse Identification of Nonlinear Dynamical Systems via $L_\infty$ Approximation}


\author{Yuqiang Wu}
\email[]{wuyuqiang@hust.edu.cn}
\affiliation{School of Artificial Intelligence and Automation, Huazhong University of Science and Technology}


\date{\today}

\begin{abstract}
This paper deals with the error processing problem of sparse identification of nonlinear dynamical systems(SINDy)
through introducing the $L_\infty$ approximation to take place of the former $L_2$ approximation.
The motivation is that the $L_\infty$ approximation could better describe the error phenomenon in the SINDy,
which consists of the derivative approximation error and the measurement noise.
Then,
an iterative thresholding algorithm is proposed to solve the reformulated problem.
3 scenarios of possible errors are considered in the experiment.
The results show that the $L_\infty$ approximation performs better or at least equal than the $L_2$ approximation in face of different error cases.
Hence,
it is reasonable to consider the $L_\infty$ approximation in the applications of the SINDy.
\end{abstract}


\maketitle

\section{Introduction}
\par Distilling information
from the observed data of an unknown system,
is an important study topic in physics\cite{bongard2007automated,schmidt2009distilling}
and many other disciplines\cite{villaverde2014reverse,martin2018reverse}.
With different kinds of data and models,
lots of mathematical modeling methods originating from the data science society have been employed,
such as statistical inference\cite{huang2009survey}
and deep learning\cite{kutz2017deep}.
In the most of the related works,
such methods are validated to provide good identification models in both fitting and prediction tasks.
Although it is very helpful to learn the input-output relationships of the unknown system,
the identified models are mathematically black-box models
which have no insight into the physical mechanisms of the real system.
\par Towards the better interpretability of the identified models,
the sparse identification of nonlinear dynamical systems(SINDy) method
is presented\cite{brunton2016discovering}
which aims to find the parsimonious mathematical equations of the unknown system from a big dictionary of the potential dynamics.
For example,
it successfully discovers the Lorenz dynamics from a dictionary consisting of the polynomial dynamics\cite{brunton2016discovering}.
The technical steps of the SINDy is explained as follows.
Firstly, the system state data is obtained through observations
and the derivatives are approximated by the numerical difference methods.
Secondly, a large candidate dictionary containing plenty of possible terms of the system dynamics
is constructed according to the prior professional information.
Thirdly, the sparse regression is performed to select a few important terms from the dictionary
to give a simple mathematical representation of the unknown system.
The SINDy is farily simple and efficient.
As a consequence, it soon gains general interests and is successfully applied to discover
a wide range of systems like the ordinary differential equations(ODEs) and the partial differential equations(PDEs)\cite{quade2018sparse,bramburger2020poincare,kaheman2020sindy}.
\par Though the SINDy is very effective to reveal governing equations,
it has two weak points in the framework,
which are illustrated in the following.
Firstly, the derivatives of the system state are calculated through the numerical difference methods.
The accuracy of this approximation has a big impact on the results of the sparse regression.
Secondly,
since the measurement noise is included in both the derivatives terms and the dictionary terms,
it also has a great influence on the performance of the algorithm.
Hence,
the error processing problem of the SINDy becomes a very important task.
To deal with the two kinds of errors mentioned above,
the most straightforward way is to modify the quality of data.
A lot of works have been presented in this aspect\cite{rudy2019deep,rudy2019smoothing,kaheman2020automatic,van2020numerical}.
These methods effectively improve the accuracy of the data and further the performance of the SINDy.
However, the improvement is not reasonable because
modifying data is independent with the SINDy and follows other principles\cite{van2020numerical}.
It is still unknown whether the modified data is closer to the true value or not.
Obviously,
if the data is over-modified,
the identification results are also incorrected.
Apart from the improvement of data,
there are methods which make adjustments to the identification framework,
which are the weak-SINDy methods\cite{schaeffer2017sparse,messenger2021weak}. 
In these works, the integral form of the governing equations is considered,
which directly avoid the derivative approximation.
In the meantime,
the impact of the measurement noise decreases through the integration.
It is rather a good way to deal with the error processing problem of the SINDy.
However,
the catch is, the integral form is less informative.
The identified model of the weak-SINDy is also weaker than the SINDy.
In summary,
the error processing problem of the SINDy, referring to the derivative approximation error and the measurement noise,
is still an open problem.
\par This paper thouroughly discusses the error processing problem of the SINDy.
The contributions of this paper are as follows.\\
\textbf{1)} The explicit error model of the sparse regression is established which explicitly describes both the derivative approximation error
and the measurement noise.\\
\textbf{2)} The $L_\infty$ approximation is introduced to take place of the $L_2$ approximation in the explicit error model.
The motivation is that
the $L_\infty$ approximation bounds the value of the residual (or the fitting error) in an interval
without considering its probability distribution.
In real world, the prior knowledge of the residual is naturally lacked.
That is, the probability distribution of the residual is unknown.
Hence,
it is intuitively better to use the $L_\infty$ approximation rather than the $L_2$ approximation to model this phenomenon.\\
\textbf{3)} To solve problem of the $L_\infty$ approximation with induced sparsity,
an iterative thresholding algorithm based on linear programming is proposed.\\
\textbf{4)} Experiments validate the effectiveness of the $L_\infty$ approximation in 3 potential error scenarios.\\
\par The remainder of this paper is organized as follows.
Section II introduces the explicit error model, the $L_\infty$ approximation and the iterative thresholding algorithm.
Section III exhibits the comparative experiments of the $L_\infty$ approximation and the $L_2$ approximation 
in different potential scenarios of the error.
Section IV discusses some technical issues.
Section V concludes this paper.

\section{Problem statement}
\par The original SINDy\cite{brunton2016discovering} is briefly reviewed in the beginning.
\subsection{Original SINDy}
\par Generally, 
consider the following 1-D ordinary differential equation(ODE):
\begin{equation}
  \frac{dx}{dt}=f(x),
\end{equation}
where $x\in\mathbb{R}$ represents the system state
and $f$ means the system dynamics which is unknown.
Through sampling in the time scale $[t_1, t_n]$,
the time series of the system state $x$ is obtained
which is written as $\mathbf{x}=[x_1, x_2, \ldots, x_n]^T$.
Then, 
the time series of the state derivative is approximated by the numerical difference methods
which is written as $\widetilde{\dot{\mathbf{x}}}=[\widetilde{\dot{x}_1}, \widetilde{\dot{x}_2}, \ldots, \widetilde{\dot{x}_n}]^T$.
Next, 
the candidate dictionary of the potential system dynamics is constructed according to the prior knowledge.
For example, an $m$th order polynomial dictionary is written as follows:
\begin{equation}
  \mathbf{\Theta}=
  \left[
   \begin{array}{ccccc}
    1 & x_1 & x^2_1 & \cdots & x^m_1 \\
    1 & x_2 & x^2_2 & \cdots & x^m_2 \\
    \vdots & \vdots & \vdots & \ddots & \vdots \\
    1 & x_n & x^2_n & \cdots & x^m_n
   \end{array}
  \right].
\end{equation}
Defining the weight vector $\mathbf{\xi}=[\xi_1, \xi_2, \ldots, \xi_m]^T$,
the sparse regression problem is established as follows:
\begin{equation}
  \mathop{\arg\min}\limits_{\mathbf{\xi}}\|\widetilde{\dot{\mathbf{x}}}-\mathbf{\Theta}\mathbf{\xi}\|_2+\lambda\|\mathbf{\xi}\|_0,
\end{equation}
where $\lambda$ is the regularization factor.
The solution of Eq.(3) refers to the identified dynamics in the dictionary.
\subsection{Explicit error model}
\par Define the measurement noise $\mathbf{w}$,
then the system state $\mathbf{x}$ is rewritten as $\mathbf{x}(\mathbf{w})$
and the dictionary $\mathbf{\Theta}$ is rewritten as $\mathbf{\Theta}(\mathbf{w})$.
Define the derivative approximation error $\mathbf{v}$,
thus the state derivative $\widetilde{\mathbf{\dot{x}}}$ is rewritten as $\widetilde{\mathbf{\dot{x}}}(\mathbf{v}, \mathbf{w})$.
With the new notations,
Eq.(3) is reformulated as:
\begin{equation}
  \mathop{\arg\min}\limits_{\mathbf{\xi}}\|\widetilde{\dot{\mathbf{x}}}(\mathbf{v}, \mathbf{w})-\mathbf{\Theta}(\mathbf{w})\mathbf{\xi}\|_2+\lambda\|\mathbf{\xi}\|_0,
\end{equation}
which is named the explicit error model of the SINDy.
\subsection{$L_\infty$ approximation}
\par For a vector $\mathbf{p}\in\mathbb{R}^n$,
its $L_2$ norm and $L_\infty$ norm are defined as follows:
\begin{equation}
  \|\mathbf{p}\|_2=(x^2_1+x^2_2+\cdots+x^2_n)^\frac{1}{2},
\end{equation}
and
\begin{equation}
\|\mathbf{p}\|_\infty=\max(|x_1|,|x_2|,\ldots,|x_n|).
\end{equation}
Define the distance vector between $\widetilde{\dot{\mathbf{x}}}(\mathbf{v}, \mathbf{w})$ and $\mathbf{\Theta}(\mathbf{w})\mathbf{\xi}$,
which is the residual $\mathbf{r}$, as follows:
\begin{equation}
  \mathbf{r}=\widetilde{\dot{\mathbf{x}}}(\mathbf{v}, \mathbf{w})-\mathbf{\Theta}(\mathbf{w})\mathbf{\xi}=[r_1, r_2, \ldots, r_n]^T.
\end{equation}
The $L_2$ approximation is to minimize the $L_2$ norm of $\mathbf{r}$ as in Eq.(4)
while the $L_\infty$ approximation is to minimize the $L_\infty$ norm of $\mathbf{r}$\cite{boyd2004convex}.
\par As is known, the $L_2$ approximation leads to the normal distribution of $r$
and the $L_\infty$ approximation bounds $r$ in the interval of $[-r_{max},\ r_{max}]$ 
where $r_{max}$ is the maximum of $|r|$.
The intuitive illustration of the $L_2$ approximation and the $L_\infty$ approximation is shown in FIG. 1.
\begin{figure}
  \centering
  \subfigure[$L_2$ approximation]{\includegraphics[width=1.5in]{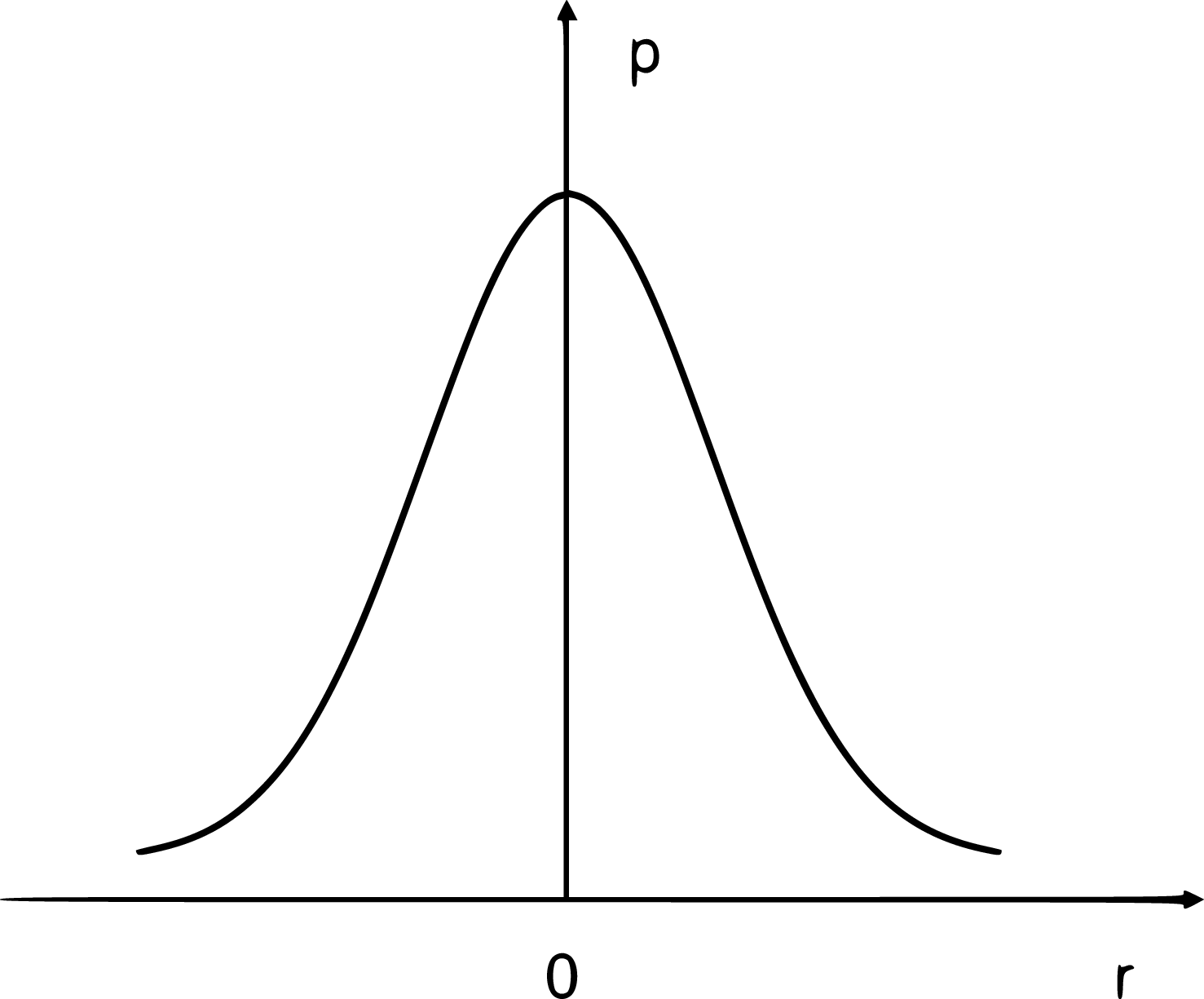}}
  \subfigure[$L_\infty$ approximation]{\includegraphics[width=1.5in]{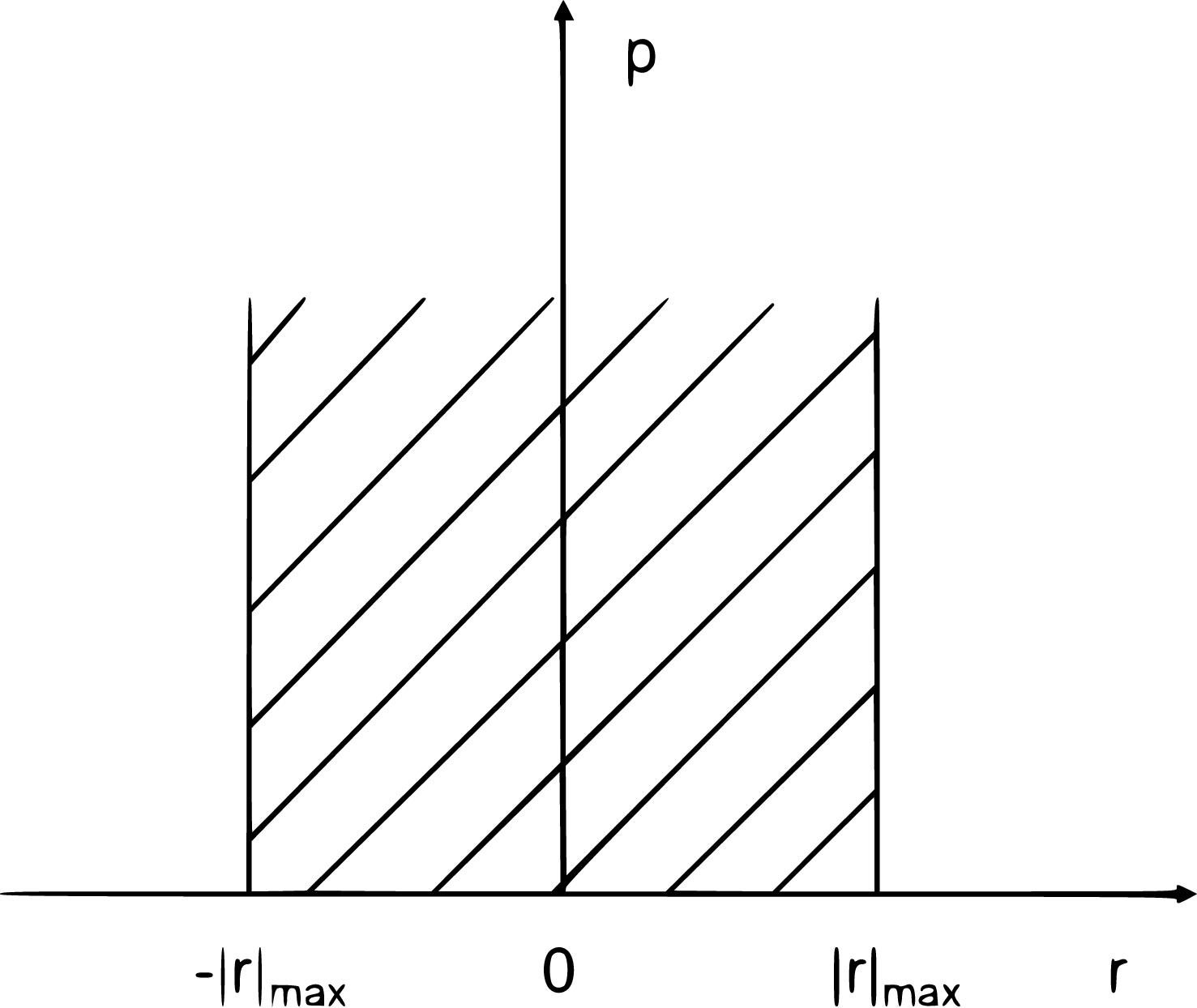}}
  \caption{Distributions of $r$. The curve in (a) represents the normal distribution and the shadow in (b) means that the distribution is unknown in this area.}
\end{figure}
Notice there is no prior knowledge of the distribution of $r$
because the distributions of the derivative approximation error $\mathbf{v}$ and the measurement noise $\mathbf{w}$ in Eq.(7)
are both unknown in general cases.
It is worth mentioning that the wide use of the $L_2$ approximation is based on the assumption that $r$ satisfies the normal distribution
but it doesn't hold in the SINDy.
On the other hand,
the $L_\infty$ approximation has no limitation on the distribution of $r$.
It is more natural to use the $L_\infty$ approximation rather than the $L_2$ approximation in Eq.(4).
Therefore,
the $L_\infty$ approximation is introduced and Eq.(4) is reformulated as follows:
\begin{equation}
  \mathop{\arg\min}\limits_{\mathbf{\xi}}\|\widetilde{\dot{\mathbf{x}}}(\mathbf{v}, \mathbf{w})-\mathbf{\Theta}(\mathbf{w})\mathbf{\xi}\|_\infty+\lambda\|\mathbf{\xi}\|_0.
\end{equation}
\subsection{Algorithm}
\par In Eq.(8),
there are the $L_\infty$ norm and the $L_0$ norm.
Without considering the $L_0$ norm,
Eq(8) can be transformed into the linear programming(LP) problem\cite{boyd2004convex}.
It can be solved by the LP solvers.
On the other hand,
the $L_0$ norm can be handled by the iterative threshoing algorithm as in SINDy\cite{brunton2016discovering,zhang2019convergence}.
Hence,
an iterative thresholding algorithm for Eq.(8) is proposed in this paper
which is shown as follows:
\begin{enumerate}[1)]
  \item $k = 0$
  \item Initialize $\mathbf{\Theta}^k$
  \item $\mathbf{\xi}^k=\mathop{\arg\min}\limits_{\mathbf{\xi}}\|\widetilde{\dot{\mathbf{x}}}(\mathbf{v}, \mathbf{w})-\mathbf{\Theta}^k(\mathbf{w})\mathbf{\xi}^k\|_\infty$ \% LP solver
  \item $k = k + 1$
  \item $\mathbf{\Theta}^k= \mathbf{\Theta}^{k-1}(|\mathbf{\xi}^{k-1}|<\mathbf{\tau)}$ \% thresholding
  \item \textbf{While} $\mathbf{\Theta}^k \neq \mathbf{\Theta}^{k-1}$
  \item \quad $\mathbf{\xi}^k=\mathop{\arg\min}\limits_{\mathbf{\xi}}\|\widetilde{\dot{\mathbf{x}}}(\mathbf{v}, \mathbf{w})-\mathbf{\Theta}^k(\mathbf{w})\mathbf{\xi}^k\|_\infty$ \% LP solver
  \item \quad $k = k + 1$
  \item \quad $\mathbf{\Theta}^k= \mathbf{\Theta}^{k-1}(|\mathbf{\xi}^{k-1}|<\mathbf{\tau)}$ \% thresholding
\item \textbf{End While}
\end{enumerate}

\section{Experiments and Results}
This section shows the simulation experiments and the results.
3 different error scenarios are considered which are explained in detail in the following.
\subsection*{Scenario 1: different sampling interval $\Delta t$ and numerical difference methods}
\par Situations without the measurement noise are considered first.
That is, only the derivative approximation error $\mathbf{v}$ is considered.
There are 2 factors that impact the accuracy of the algorithm,
which are the sampling interval $\Delta t$ and the numerical difference method.
In the experiments,
different combinations of the sampling interval $\Delta t$ and the numerical difference method
are tested.
Lorenz system is considered which is formulated as follows:
\begin{equation}
    \begin{array}{l}\left\{\begin{array}{l}\dot{x}=\sigma(y-x)\\\dot{y}=x(\rho-z)-y\\\dot{z}=xy-\beta z\end{array}\right.,\end{array}
\end{equation}
where $\sigma=10$, $\rho=28$, $\beta=8/3$.
The initial state is
$[-8, 8, 27]^T$ and the time scale is $[0, 100s]$.
Note that the last 5000 data points of the time series are adopted in the experiment to avoid the transient states.
Besides, in the reconstruction stage,
the initial state is set as $[1, 1, 1]^T$ and the sampling interval keeps the same as in the identification stage.
\par The results are shown in Table I.
Combinations of the sampling interval $\Delta t$ and the numerical difference method are listed in rows.
In each row (a single case),
the simulation errors of the $L_\infty$ approximation and the $L_2$ approximation 
with root mean square error(RMSE) and standard deviation(STD) are shown.
The error in three dimensions of Lorenz system and the total of three dimensions are compared.
The dash means in that case,
the identification fails so we cannot calculate the reconstruction result.
\par In the comparison of three numerical difference methods,
it is shown that the central difference method is the most robust one
while the forward difference method and the total-variation regularization method fail in many cases.
In the comparison of different sampling intervals,
it doesn't show the direct correlation between the error and the interval,
in both the independent dimension and the total.
In the comparison of the $L_\infty$ approximation and the $L_2$ approximation,
it shows that the reconstruction errors of two methods are close.
However,
the $L_\infty$ approximation has more fail cases than the $L_2$ approximation.
This will be discussed in section IV.

\begin{table*}
\footnotesize
\centering
\caption{Comparison results of the $L_\infty$ approximation and the $L_2$ approximation of different derivative approximation errors.}
\resizebox{\linewidth}{!}{
\begin{tabular}{c|c|cc|cc||cc|cc||cc|cc||cc|cc}
  \toprule
   \multirow{3}{2cm}{Numerical method}&\multirow{3}{*}{$\Delta t$}&\multicolumn{4}{c||}{Dimension $x$}&\multicolumn{4}{c||}{Dimension $y$}&\multicolumn{4}{c||}{Dimension $z$}&\multicolumn{4}{c}{Total}\\
 & & \multicolumn{2}{c|}{RMSE} & \multicolumn{2}{c||}{STD}& \multicolumn{2}{c|}{RMSE} & \multicolumn{2}{c||}{STD}& \multicolumn{2}{c|}{RMSE} & \multicolumn{2}{c||}{STD}& \multicolumn{2}{c|}{RMSE} & \multicolumn{2}{c}{STD}\\
   & &$L_2$& $L_\infty$&$L_2$& $L_\infty$&$L_2$& $L_\infty$&$L_2$& $L_\infty$&$L_2$& $L_\infty$&$L_2$& $L_\infty$&$L_2$& $L_\infty$&$L_2$& $L_\infty$\\\midrule
\multirow{7}{2cm}{Central difference}&0.001&11.6728&13.4214&11.6308&11.5605&13.2821&14.9268&13.2329&13.3129&15.1936&9.7550&15.1640&9.7056&13.4600&12.8854&13.4546&11.9365\\
&0.0025&9.1804&--&8.9107&--&10.9089&--&10.6615&--&11.0812&--&11.0620&--&10.4256&--&10.3463&--\\
&0.005&10.7893&10.8471&10.7090&10.7846&11.9710&12.4904&11.8994&12.4304&11.5118&13.1615&11.4266&13.1590&11.4344&12.2051&11.3548&12.1717\\
&0.0075&11.5706&11.4161&11.4682&11.1201&13.1704&12.8256&13.0867&12.5748&10.6479&12.2908&10.6490&12.2901&11.8422&12.1914&11.7997&12.0615\\
&0.01&11.3000&11.3670&11.2475&11.2985&12.8031&12.9600&12.7512&12.8965&13.9649&11.3513&13.9638&11.3524&12.7362&11.9167&12.7085&11.8871\\
&0.015&11.3914&11.3902&11.3122&11.2737&12.9389&12.8670&12.8699&12.7667&12.0827&13.4687&12.0815&13.4699&12.1541&12.6056&12.1269&12.5577\\
&0.02&10.8183&10.2520&10.5766&10.1911&12.2605&11.7496&12.0443&11.6952&11.5557&11.8612&11.5566&11.8622&11.5599&11.3114&11.4553&11.2882\\\midrule
\multirow{5}{2cm}{Forward difference}&0.001&10.9049&--&9.8965&--&12.4051&--&11.6640&--&14.7304&--&14.6524&--&12.7774&--&12.3045&--\\
&0.0025&10.7025&9.9047&10.5011&7.3294&12.0592&10.9949&11.7612&8.5470&7.3444&9.7468&7.2892&8.7719&10.2292&10.2306&10.1492&9.7467\\
&0.005&9.8630&--&7.3471&--&10.9503&--&8.5777&--&9.7861&--&8.7997&--&10.2137&--&9.7526&--\\
&0.0075&--&--&--&--&--&--&--&--&--&--&--&--&--&--&--&--\\
&0.01&--&11.1453&--&7.8656&--&12.1514&--&8.9947&--&9.3639&--&8.7367&--&10.9477&--&10.0969\\
&0.015&10.9150&11.2541&7.8551&7.8543&11.8636&12.4437&8.9756&8.9744&9.3479&9.1960&8.6896&8.6863&10.7589&11.0464&10.0027&10.0645\\
&0.02&9.8624&--&7.6812&--&10.8191&--&8.6999&--&8.7851&--&8.2868&--&9.8573&--&9.3122&--\\\midrule
\multirow{5}{2cm}{Total-variation regularization}&0.001&10.6920&--&10.4561&--&12.2173&--&12.0903&--&16.1909&--&16.1793&--&13.2379&--&13.1470&--\\
&0.0025&9.7147&11.0512&7.5850&10.5737&10.9473&12.2943&8.8397&11.8751&9.5156&12.4294&8.8365&12.3178&10.0791&11.9411&9.6229&11.6341\\
&0.005&12.4016&--&11.7224&--&14.1170&--&13.5913&--&6.5111&--&6.5097&--&11.4816&--&11.1653&--\\
&0.0075&--&--&--&--&--&--&--&--&--&--&--&--&--&--&--&--\\
&0.01&11.5916&--&11.5497&--&13.5597&--&13.5420&--&9.9561&--&9.8301&--&11.7948&--&11.7437&--\\
&0.015&--&--&--&--&--&--&--&--&--&--&--&--&--&--&--&--\\
&0.02&--&--&--&--&--&--&--&--&--&--&--&--&--&--&--&--\\
  \bottomrule
\end{tabular}}
\end{table*}
\subsection*{Scenario 2: different distributions of the measurement noise}
\par Measurement noise appears in the most of the natural and engineering systems.
Usually,
the measurement noise $w$ is considered as the additive Gaussian distribution $\mathcal{N}(\mu, \sigma^2)$.
However,
in general cases,
the noise distribution is unknown.
Therefore,
we test the senario with different distributions of the measurement noise.
The basic experiemnt settings are the same as in scenario 1.
Note that the central difference method is adopted here because it is the most roboust method according to senario 1.
In addition, for the convenience of comparison, the sampling interval keeps 0.01s in both
the identification stage and the reconstruction stage.
Due to the random test,
the experiment is independently run 20 times for each cese
and the statistical results are collected.
\par Table II shows the results of different measurement noise.
The error here stands for the total error of 3 dimensions.
We test 4 probability distributions, which are the normal distribution,
the uniform distribution,
the Weibull distribution
and the extreme value distribution.
In each distribution,
different parameter settings are considered to cover the possible cases of the distribution.
For example,
in the normal distribution,
$\sigma$ is fixed in $0.01$ and $\mu$ is changed from $-0.02$ to $0.02$ with the step interval of $0.01$,
in order to make different skew distributions.
To emphasize,
the purpose to make different distributions is to simulate the measurement noise in the real environment
because it is hard to know the noise distribution.
\par In Table II,
the bold nunber is the better one in the comparison of the $L_\infty$ approximation and the $L_2$ approximation.
Note that the better is related to the mean of RMSE or STD in the statistical view rather than the terms in the brackets.
It is shown that the $L_\infty$ approximation has a better performance in most cases even under the normal distribution.
\begin{table*}
\footnotesize
\centering
\caption{Results with different distributions. The bold number is the better one in the comparison of the $L_\infty$ approximation and the $L_2$ approximation.}
\begin{tabular}{c|c|cc|cc}
  \toprule
\multirow{2}{*}{Distributions}&\multirow{2}{*}{Parameters}&\multicolumn{2}{c|}{RMSE} & \multicolumn{2}{c}{STD}\\
 & &$L_2$& $L_\infty$&$L_2$& $L_\infty$\\\midrule
\multirow{5}{*}{Normal}&$\mu=-0.02,\sigma=0.01$&$12.2789(\pm0.4769)$&$\mathbf{12.1505}(\pm0.6636)$&$12.2455(\pm0.4802)$&$\mathbf{12.1270}(\pm0.6614)$\\
&$\mu=-0.01,\sigma=0.01$&$12.0622(\pm0.6171)$&$\mathbf{12.0005}(\pm0.4913)$&$12.0173(\pm0.6073)$&$\mathbf{11.9501}(\pm0.4984)$\\
&$\mu=0,\sigma=0.01$&$11.9794(\pm0.4834)$&$\mathbf{11.9202}(\pm0.5028)$&$11.9386(\pm0.4804)$&$\mathbf{11.8955}(\pm0.5092)$\\
&$\mu=0.01,\sigma=0.01$&$12.0948(\pm0.4375)$&$\mathbf{11.7917}(\pm0.6565)$&$12.0576(\pm0.4435)$&$\mathbf{11.7600}(\pm0.6450)$\\
&$\mu=0.02,\sigma=0.01$&$\mathbf{11.8617}(\pm0.5514)$&$12.0592(\pm0.4613)$&$\mathbf{11.8063}(\pm0.5501)$&$12.0085(\pm0.4714)$\\\midrule
\multirow{3}{*}{Uniform}&$a=-0.02, b=0$&$12.0739(\pm0.5702)$&$\mathbf{12.0186}(\pm0.4622)$&$12.0308(\pm0.5865)$&$\mathbf{11.9763}(\pm0.4527)$\\
&$a=-0.01, b=0.01$&$12.1744(\pm0.5948)$&$\mathbf{12.1144}(\pm0.6095)$&$12.1239(\pm0.6068)$&$\mathbf{12.0726}(\pm0.6182)$\\
&$a=0, b=0.02$&$\mathbf{11.9169}(\pm0.5870)$&$12.1381(\pm0.6329)$&$\mathbf{11.8844}(\pm0.5753)$&$12.0950(\pm0.6491)$\\\midrule
\multirow{2}{*}{Weibull}&$a=0.01, b=4$&$12.1127(\pm0.6207)$&$\mathbf{11.8376}(\pm0.5438)$&$12.0704(\pm0.6034)$&$\mathbf{11.8161}(\pm0.5416)$\\
&$a=0.02, b=4$&$12.1908(\pm0.5707)$&$\mathbf{12.1231}(\pm0.5122)$&$12.1403(\pm0.5871)$&$\mathbf{12.1019}(\pm0.5149)$\\\midrule
\multirow{3}{*}{Extreme value}&$\mu=-0.01, \sigma=0.01$&$12.0490(\pm0.4765)$&$\mathbf{11.8725}(\pm0.6232)$&$12.0185(\pm0.4623)$&$\mathbf{11.8352}(\pm0.6233)$\\
&$\mu=0, \sigma=0.01$&$\mathbf{11.8436}(\pm0.6752)$&$11.8639(\pm0.6719)$&$\mathbf{11.8002}(\pm0.6743)$&$11.8293(\pm0.6852)$\\
&$\mu=0.01, \sigma=0.01$&$\mathbf{11.8574}(\pm0.6602)$&$12.0527(\pm0.4169)$&$\mathbf{11.8099}(\pm0.6592)$&$12.0214(\pm0.4104)$\\
\bottomrule
\end{tabular}
\end{table*}
\subsection*{Scenario 3: different combinations of $\mathbf{v}$ and $\mathbf{w}$}
\par After testing the separate influence of the derivative approximation error $\mathbf{v}$ and the measurement noise $\mathbf{w}$ in the
above experiments,
the senario with different combinations of $\mathbf{v}$ and $\mathbf{w}$ is studied here.
The expriment setting is slightly different with the former
where the number of the data points used in both the identification stage and the reconstruction stage is set 1000
because the sampling intervals in this senario become bigger than the previous.
It is worth mentioning that 
considering the wide existence and the importance of the Gaussian noise,
only different cases with the normal distribution are focused in the experiment.
\par Further, the comparison experiment of the $L_\infty$ approximation and the $L_2$ approximation are also performed
in Chen system,
which is formulated as:
\begin{equation}
    \begin{array}{l}\left\{\begin{array}{l}\dot{x}=a(y-x)\\\dot{y}=(c-a)x+cy-xz\\\dot{z}=xy-bz\end{array}\right.,\end{array}
\end{equation}
where $a=35$, $b=3$, and $c=28$.
\par The results are listed in Tablel III and Table IV, which are respectively related to Lorenz system and Chen system.
The bold number means the better one in the comparison of the $L_\infty$ approximation and the $L_2$ approximation.
It is shown that in Lorenz system,
the $L_\infty$ approximation has a better performance in most cases.
While in Chen system,
the $L_\infty$ approximation and the $L_2$ approximation have a close performance.
\begin{table*}
\footnotesize
\centering
\caption{Results of Lorenz system in scenario 3. The bold number means the better one in the comparison of the $L_\infty$ approximation and the $L_2$ approximation.}
\begin{tabular}{c|c|cc|cc}
  \toprule
\multirow{2}{*}{$\Delta t$}&\multirow{2}{*}{Parameters}&\multicolumn{2}{c|}{RMSE} & \multicolumn{2}{c}{STD}\\
 & &$L_2$& $L_\infty$&$L_2$& $L_\infty$\\\midrule
\multirow{3}{*}{0.01}&$\mu=0,\sigma=0.01$&$12.0279(\pm0.8177)$&$\mathbf{11.8673}(\pm0.8531)$&$11.8055(\pm0.7566)$&$\mathbf{11.5990}(\pm0.7826)$\\
&$\mu=0,\sigma=0.02$&$12.1645(\pm0.8010)$&$\mathbf{11.6123}(\pm1.1389)$&$11.9143(\pm0.8346)$&$\mathbf{11.3056}(\pm1.0039)$\\
&$\mu=0,\sigma=0.03$&$12.1154(\pm0.7986)$&$\mathbf{11.9702}(\pm0.6099)$&$11.9562(\pm0.7902)$&$\mathbf{11.7664}(\pm0.6002)$\\\midrule
\multirow{3}{*}{0.02}&$\mu=0,\sigma=0.01$&$12.0732(\pm0.4365)$&$\mathbf{12.0078}(\pm0.8221)$&$11.9364(\pm0.4417)$&$\mathbf{11.9281}(\pm0.8433)$\\
&$\mu=0,\sigma=0.02$&$\mathbf{11.9023}(\pm0.6838)$&$12.0773(\pm0.6521)$&$\mathbf{11.8211}(\pm0.6433)$&$11.8812(\pm0.6348)$\\
&$\mu=0,\sigma=0.03$&$12.0612(\pm0.6256)$&$\mathbf{11.8345}(\pm0.8893)$&$11.9348(\pm0.6700)$&$\mathbf{11.6734}(\pm0.9054)$\\\midrule
\bottomrule
\end{tabular}
\end{table*}
\begin{table*}
\footnotesize
\centering
\caption{Results of Chen system in scenario 3. The bold number means the better one in the comparison of the $L_\infty$ approximation and the $L_2$ approximation.}
\begin{tabular}{c|c|cc|cc}
  \toprule
\multirow{2}{*}{$\Delta t$}&\multirow{2}{*}{Parameters}&\multicolumn{2}{c|}{RMSE} & \multicolumn{2}{c}{STD}\\
 & &$L_2$& $L_\infty$&$L_2$& $L_\infty$\\\midrule
\multirow{3}{*}{0.01}&$\mu=0,\sigma=0.01$&$11.3388(\pm1.2708)$&$\mathbf{11.3040}(\pm1.0657)$&$11.3113(\pm1.2619)$&$\mathbf{11.2826}(\pm1.0652)$\\
&$\mu=0,\sigma=0.02$&$\mathbf{10.9406}(\pm0.9185)$&$11.1858(\pm0.7536)$&$\mathbf{10.9086}(\pm0.9189)$&$11.1616(\pm0.7569)$\\
&$\mu=0,\sigma=0.03$&$11.2830(\pm1.1880)$&$\mathbf{11.0228}(\pm1.0531)$&$11.2537(\pm1.2029)$&$\mathbf{11.0037}(\pm1.0565)$\\\midrule
\multirow{3}{*}{0.02}&$\mu=0,\sigma=0.01$&$11.0319(\pm0.7164)$&$\mathbf{10.7523}(\pm0.4636)$&$11.0301(\pm0.7170)$&$\mathbf{10.7447}(\pm0.4628)$\\
&$\mu=0,\sigma=0.02$&$\mathbf{10.7941}(\pm0.9526)$&$10.8860(\pm0.5015)$&$\mathbf{10.7931}(\pm0.9534)$&$10.8769(\pm0.4994)$\\
&$\mu=0,\sigma=0.03$&$\mathbf{10.7758}(\pm0.8668)$&$10.9523(\pm0.4468)$&$\mathbf{10.7745}(\pm0.8680)$&$10.9435(\pm0.4503)$\\\midrule
\bottomrule
\end{tabular}
\end{table*}
\subsection*{Summary}
\par Above experiments consider different possible scenarios of the derivative approximation error $\mathbf{v}$
and the measurement noise $\mathbf{w}$.
According to the results, 
the performance of the $L_\infty$ approximation is better or at least equal to the $L_2$ approximation.
\section{Discussions}
\par
This section talks about other issues of the $L_\infty$ approximation besides the performance in the simulation error.\\
\textbf{1)} The proposed algorithm is less efficient than the original SINDy.
The reason is that Eq.(8) is a linear programming (LP) problem while the least square (LS) method is used in the original SINDy.
The efficiency of our algorithm is strongly related to the problem scale,
which refers to the length of the time series.
If we use more data points,
the problem scale increases and the efficiency of the algorithm decreases.\\
\textbf{2)} The proposed algorithm is less robust than the original SINDy.
Although our algorithm has a better performance in the simulation error,
it is more likely to fail when Eq.(8) becomes more complex,
mainly referring to the complex constraints in LP.
\section{Conclusions}
\par This paper deals with the error processing problem in the SINDy.
The $L_\infty$ approximation is introduced to take place of the former the $L_2$ approximation.
The experimental results indicate that the $L_\infty$ approximation is an 
effective way to obtain good identification results in face of different error cases
and perform even better than the $L_2$ approximation.
Hence,
it is reasonale to consider the $L_\infty$ approximation as an alternative of the $L_2$ approximation to deal with
the error processing problem of the SINDy.

\begin{acknowledgments}
This work is supported by the National Key R\&D Program of China [Grant number 2018YFB1701202].
\end{acknowledgments}

\bibliography{apstemplatebib}

\begin{thebibliography}{18}%
\makeatletter
\providecommand \@ifxundefined [1]{%
 \@ifx{#1\undefined}
}%
\providecommand \@ifnum [1]{%
 \ifnum #1\expandafter \@firstoftwo
 \else \expandafter \@secondoftwo
 \fi
}%
\providecommand \@ifx [1]{%
 \ifx #1\expandafter \@firstoftwo
 \else \expandafter \@secondoftwo
 \fi
}%
\providecommand \natexlab [1]{#1}%
\providecommand \enquote  [1]{``#1''}%
\providecommand \bibnamefont  [1]{#1}%
\providecommand \bibfnamefont [1]{#1}%
\providecommand \citenamefont [1]{#1}%
\providecommand \href@noop [0]{\@secondoftwo}%
\providecommand \href [0]{\begingroup \@sanitize@url \@href}%
\providecommand \@href[1]{\@@startlink{#1}\@@href}%
\providecommand \@@href[1]{\endgroup#1\@@endlink}%
\providecommand \@sanitize@url [0]{\catcode `\\12\catcode `\$12\catcode
  `\&12\catcode `\#12\catcode `\^12\catcode `\_12\catcode `\%12\relax}%
\providecommand \@@startlink[1]{}%
\providecommand \@@endlink[0]{}%
\providecommand \url  [0]{\begingroup\@sanitize@url \@url }%
\providecommand \@url [1]{\endgroup\@href {#1}{\urlprefix }}%
\providecommand \urlprefix  [0]{URL }%
\providecommand \Eprint [0]{\href }%
\providecommand \doibase [0]{https://doi.org/}%
\providecommand \selectlanguage [0]{\@gobble}%
\providecommand \bibinfo  [0]{\@secondoftwo}%
\providecommand \bibfield  [0]{\@secondoftwo}%
\providecommand \translation [1]{[#1]}%
\providecommand \BibitemOpen [0]{}%
\providecommand \bibitemStop [0]{}%
\providecommand \bibitemNoStop [0]{.\EOS\space}%
\providecommand \EOS [0]{\spacefactor3000\relax}%
\providecommand \BibitemShut  [1]{\csname bibitem#1\endcsname}%
\let\auto@bib@innerbib\@empty
\bibitem [{\citenamefont {Bongard}\ and\ \citenamefont
  {Lipson}(2007)}]{bongard2007automated}%
  \BibitemOpen
  \bibfield  {author} {\bibinfo {author} {\bibfnamefont {J.}~\bibnamefont
  {Bongard}}\ and\ \bibinfo {author} {\bibfnamefont {H.}~\bibnamefont
  {Lipson}},\ }\bibfield  {title} {\bibinfo {title} {Automated reverse
  engineering of nonlinear dynamical systems},\ }\href@noop {} {\bibfield
  {journal} {\bibinfo  {journal} {Proceedings of the National Academy of
  Sciences}\ }\textbf {\bibinfo {volume} {104}},\ \bibinfo {pages} {9943}
  (\bibinfo {year} {2007})}\BibitemShut {NoStop}%
\bibitem [{\citenamefont {Schmidt}\ and\ \citenamefont
  {Lipson}(2009)}]{schmidt2009distilling}%
  \BibitemOpen
  \bibfield  {author} {\bibinfo {author} {\bibfnamefont {M.}~\bibnamefont
  {Schmidt}}\ and\ \bibinfo {author} {\bibfnamefont {H.}~\bibnamefont
  {Lipson}},\ }\bibfield  {title} {\bibinfo {title} {Distilling free-form
  natural laws from experimental data},\ }\href@noop {} {\bibfield  {journal}
  {\bibinfo  {journal} {science}\ }\textbf {\bibinfo {volume} {324}},\ \bibinfo
  {pages} {81} (\bibinfo {year} {2009})}\BibitemShut {NoStop}%
\bibitem [{\citenamefont {Villaverde}\ and\ \citenamefont
  {Banga}(2014)}]{villaverde2014reverse}%
  \BibitemOpen
  \bibfield  {author} {\bibinfo {author} {\bibfnamefont {A.~F.}\ \bibnamefont
  {Villaverde}}\ and\ \bibinfo {author} {\bibfnamefont {J.~R.}\ \bibnamefont
  {Banga}},\ }\bibfield  {title} {\bibinfo {title} {Reverse engineering and
  identification in systems biology: strategies, perspectives and challenges},\
  }\href@noop {} {\bibfield  {journal} {\bibinfo  {journal} {Journal of the
  Royal Society Interface}\ }\textbf {\bibinfo {volume} {11}},\ \bibinfo
  {pages} {20130505} (\bibinfo {year} {2014})}\BibitemShut {NoStop}%
\bibitem [{\citenamefont {Martin}\ \emph {et~al.}(2018)\citenamefont {Martin},
  \citenamefont {Munch},\ and\ \citenamefont {Hein}}]{martin2018reverse}%
  \BibitemOpen
  \bibfield  {author} {\bibinfo {author} {\bibfnamefont {B.~T.}\ \bibnamefont
  {Martin}}, \bibinfo {author} {\bibfnamefont {S.~B.}\ \bibnamefont {Munch}},\
  and\ \bibinfo {author} {\bibfnamefont {A.~M.}\ \bibnamefont {Hein}},\
  }\bibfield  {title} {\bibinfo {title} {Reverse-engineering ecological theory
  from data},\ }\href@noop {} {\bibfield  {journal} {\bibinfo  {journal}
  {Proceedings of the Royal Society B: Biological Sciences}\ }\textbf {\bibinfo
  {volume} {285}},\ \bibinfo {pages} {20180422} (\bibinfo {year}
  {2018})}\BibitemShut {NoStop}%
\bibitem [{\citenamefont {Huang}\ \emph {et~al.}(2009)\citenamefont {Huang},
  \citenamefont {Tienda-Luna},\ and\ \citenamefont {Wang}}]{huang2009survey}%
  \BibitemOpen
  \bibfield  {author} {\bibinfo {author} {\bibfnamefont {Y.}~\bibnamefont
  {Huang}}, \bibinfo {author} {\bibfnamefont {I.~M.}\ \bibnamefont
  {Tienda-Luna}},\ and\ \bibinfo {author} {\bibfnamefont {Y.}~\bibnamefont
  {Wang}},\ }\bibfield  {title} {\bibinfo {title} {A survey of statistical
  models for reverse engineering gene regulatory networks},\ }\href@noop {}
  {\bibfield  {journal} {\bibinfo  {journal} {IEEE signal processing magazine}\
  }\textbf {\bibinfo {volume} {26}},\ \bibinfo {pages} {76} (\bibinfo {year}
  {2009})}\BibitemShut {NoStop}%
\bibitem [{\citenamefont {Kutz}(2017)}]{kutz2017deep}%
  \BibitemOpen
  \bibfield  {author} {\bibinfo {author} {\bibfnamefont {J.~N.}\ \bibnamefont
  {Kutz}},\ }\bibfield  {title} {\bibinfo {title} {Deep learning in fluid
  dynamics},\ }\href@noop {} {\bibfield  {journal} {\bibinfo  {journal}
  {Journal of Fluid Mechanics}\ }\textbf {\bibinfo {volume} {814}},\ \bibinfo
  {pages} {1} (\bibinfo {year} {2017})}\BibitemShut {NoStop}%
\bibitem [{\citenamefont {Brunton}\ \emph {et~al.}(2016)\citenamefont
  {Brunton}, \citenamefont {Proctor},\ and\ \citenamefont
  {Kutz}}]{brunton2016discovering}%
  \BibitemOpen
  \bibfield  {author} {\bibinfo {author} {\bibfnamefont {S.~L.}\ \bibnamefont
  {Brunton}}, \bibinfo {author} {\bibfnamefont {J.~L.}\ \bibnamefont
  {Proctor}},\ and\ \bibinfo {author} {\bibfnamefont {J.~N.}\ \bibnamefont
  {Kutz}},\ }\bibfield  {title} {\bibinfo {title} {Discovering governing
  equations from data by sparse identification of nonlinear dynamical
  systems},\ }\href@noop {} {\bibfield  {journal} {\bibinfo  {journal}
  {Proceedings of the national academy of sciences}\ }\textbf {\bibinfo
  {volume} {113}},\ \bibinfo {pages} {3932} (\bibinfo {year}
  {2016})}\BibitemShut {NoStop}%
\bibitem [{\citenamefont {Quade}\ \emph {et~al.}(2018)\citenamefont {Quade},
  \citenamefont {Abel}, \citenamefont {Nathan~Kutz},\ and\ \citenamefont
  {Brunton}}]{quade2018sparse}%
  \BibitemOpen
  \bibfield  {author} {\bibinfo {author} {\bibfnamefont {M.}~\bibnamefont
  {Quade}}, \bibinfo {author} {\bibfnamefont {M.}~\bibnamefont {Abel}},
  \bibinfo {author} {\bibfnamefont {J.}~\bibnamefont {Nathan~Kutz}},\ and\
  \bibinfo {author} {\bibfnamefont {S.~L.}\ \bibnamefont {Brunton}},\
  }\bibfield  {title} {\bibinfo {title} {Sparse identification of nonlinear
  dynamics for rapid model recovery},\ }\href@noop {} {\bibfield  {journal}
  {\bibinfo  {journal} {Chaos: An Interdisciplinary Journal of Nonlinear
  Science}\ }\textbf {\bibinfo {volume} {28}},\ \bibinfo {pages} {063116}
  (\bibinfo {year} {2018})}\BibitemShut {NoStop}%
\bibitem [{\citenamefont {Bramburger}\ and\ \citenamefont
  {Kutz}(2020)}]{bramburger2020poincare}%
  \BibitemOpen
  \bibfield  {author} {\bibinfo {author} {\bibfnamefont {J.~J.}\ \bibnamefont
  {Bramburger}}\ and\ \bibinfo {author} {\bibfnamefont {J.~N.}\ \bibnamefont
  {Kutz}},\ }\bibfield  {title} {\bibinfo {title} {Poincar{\'e} maps for
  multiscale physics discovery and nonlinear floquet theory},\ }\href@noop {}
  {\bibfield  {journal} {\bibinfo  {journal} {Physica D: Nonlinear Phenomena}\
  }\textbf {\bibinfo {volume} {408}},\ \bibinfo {pages} {132479} (\bibinfo
  {year} {2020})}\BibitemShut {NoStop}%
\bibitem [{\citenamefont {Kaheman}\ \emph
  {et~al.}(2020{\natexlab{a}})\citenamefont {Kaheman}, \citenamefont {Kutz},\
  and\ \citenamefont {Brunton}}]{kaheman2020sindy}%
  \BibitemOpen
  \bibfield  {author} {\bibinfo {author} {\bibfnamefont {K.}~\bibnamefont
  {Kaheman}}, \bibinfo {author} {\bibfnamefont {J.~N.}\ \bibnamefont {Kutz}},\
  and\ \bibinfo {author} {\bibfnamefont {S.~L.}\ \bibnamefont {Brunton}},\
  }\bibfield  {title} {\bibinfo {title} {Sindy-pi: a robust algorithm for
  parallel implicit sparse identification of nonlinear dynamics},\ }\href@noop
  {} {\bibfield  {journal} {\bibinfo  {journal} {Proceedings of the Royal
  Society A}\ }\textbf {\bibinfo {volume} {476}},\ \bibinfo {pages} {20200279}
  (\bibinfo {year} {2020}{\natexlab{a}})}\BibitemShut {NoStop}%
\bibitem [{\citenamefont {Rudy}\ \emph
  {et~al.}(2019{\natexlab{a}})\citenamefont {Rudy}, \citenamefont {Kutz},\ and\
  \citenamefont {Brunton}}]{rudy2019deep}%
  \BibitemOpen
  \bibfield  {author} {\bibinfo {author} {\bibfnamefont {S.~H.}\ \bibnamefont
  {Rudy}}, \bibinfo {author} {\bibfnamefont {J.~N.}\ \bibnamefont {Kutz}},\
  and\ \bibinfo {author} {\bibfnamefont {S.~L.}\ \bibnamefont {Brunton}},\
  }\bibfield  {title} {\bibinfo {title} {Deep learning of dynamics and
  signal-noise decomposition with time-stepping constraints},\ }\href@noop {}
  {\bibfield  {journal} {\bibinfo  {journal} {Journal of Computational
  Physics}\ }\textbf {\bibinfo {volume} {396}},\ \bibinfo {pages} {483}
  (\bibinfo {year} {2019}{\natexlab{a}})}\BibitemShut {NoStop}%
\bibitem [{\citenamefont {Rudy}\ \emph
  {et~al.}(2019{\natexlab{b}})\citenamefont {Rudy}, \citenamefont {Brunton},\
  and\ \citenamefont {Kutz}}]{rudy2019smoothing}%
  \BibitemOpen
  \bibfield  {author} {\bibinfo {author} {\bibfnamefont {S.~H.}\ \bibnamefont
  {Rudy}}, \bibinfo {author} {\bibfnamefont {S.~L.}\ \bibnamefont {Brunton}},\
  and\ \bibinfo {author} {\bibfnamefont {J.~N.}\ \bibnamefont {Kutz}},\
  }\bibfield  {title} {\bibinfo {title} {Smoothing and parameter estimation by
  soft-adherence to governing equations},\ }\href@noop {} {\bibfield  {journal}
  {\bibinfo  {journal} {Journal of Computational Physics}\ }\textbf {\bibinfo
  {volume} {398}},\ \bibinfo {pages} {108860} (\bibinfo {year}
  {2019}{\natexlab{b}})}\BibitemShut {NoStop}%
\bibitem [{\citenamefont {Kaheman}\ \emph
  {et~al.}(2020{\natexlab{b}})\citenamefont {Kaheman}, \citenamefont
  {Brunton},\ and\ \citenamefont {Kutz}}]{kaheman2020automatic}%
  \BibitemOpen
  \bibfield  {author} {\bibinfo {author} {\bibfnamefont {K.}~\bibnamefont
  {Kaheman}}, \bibinfo {author} {\bibfnamefont {S.~L.}\ \bibnamefont
  {Brunton}},\ and\ \bibinfo {author} {\bibfnamefont {J.~N.}\ \bibnamefont
  {Kutz}},\ }\bibfield  {title} {\bibinfo {title} {Automatic differentiation to
  simultaneously identify nonlinear dynamics and extract noise probability
  distributions from data},\ }\href@noop {} {\bibfield  {journal} {\bibinfo
  {journal} {arXiv preprint arXiv:2009.08810}\ } (\bibinfo {year}
  {2020}{\natexlab{b}})}\BibitemShut {NoStop}%
\bibitem [{\citenamefont {van Breugel}\ \emph {et~al.}(2020)\citenamefont {van
  Breugel}, \citenamefont {Kutz},\ and\ \citenamefont
  {Brunton}}]{van2020numerical}%
  \BibitemOpen
  \bibfield  {author} {\bibinfo {author} {\bibfnamefont {F.}~\bibnamefont {van
  Breugel}}, \bibinfo {author} {\bibfnamefont {J.~N.}\ \bibnamefont {Kutz}},\
  and\ \bibinfo {author} {\bibfnamefont {B.~W.}\ \bibnamefont {Brunton}},\
  }\bibfield  {title} {\bibinfo {title} {Numerical differentiation of noisy
  data: A unifying multi-objective optimization framework},\ }\href@noop {}
  {\bibfield  {journal} {\bibinfo  {journal} {IEEE Access}\ } (\bibinfo {year}
  {2020})}\BibitemShut {NoStop}%
\bibitem [{\citenamefont {Schaeffer}\ and\ \citenamefont
  {McCalla}(2017)}]{schaeffer2017sparse}%
  \BibitemOpen
  \bibfield  {author} {\bibinfo {author} {\bibfnamefont {H.}~\bibnamefont
  {Schaeffer}}\ and\ \bibinfo {author} {\bibfnamefont {S.~G.}\ \bibnamefont
  {McCalla}},\ }\bibfield  {title} {\bibinfo {title} {Sparse model selection
  via integral terms},\ }\href@noop {} {\bibfield  {journal} {\bibinfo
  {journal} {Physical Review E}\ }\textbf {\bibinfo {volume} {96}},\ \bibinfo
  {pages} {023302} (\bibinfo {year} {2017})}\BibitemShut {NoStop}%
\bibitem [{\citenamefont {Messenger}\ and\ \citenamefont
  {Bortz}(2021)}]{messenger2021weak}%
  \BibitemOpen
  \bibfield  {author} {\bibinfo {author} {\bibfnamefont {D.~A.}\ \bibnamefont
  {Messenger}}\ and\ \bibinfo {author} {\bibfnamefont {D.~M.}\ \bibnamefont
  {Bortz}},\ }\bibfield  {title} {\bibinfo {title} {Weak sindy: Galerkin-based
  data-driven model selection},\ }\href@noop {} {\bibfield  {journal} {\bibinfo
   {journal} {Multiscale Modeling \& Simulation}\ }\textbf {\bibinfo {volume}
  {19}},\ \bibinfo {pages} {1474} (\bibinfo {year} {2021})}\BibitemShut
  {NoStop}%
\bibitem [{\citenamefont {Boyd}\ \emph {et~al.}(2004)\citenamefont {Boyd},
  \citenamefont {Boyd},\ and\ \citenamefont {Vandenberghe}}]{boyd2004convex}%
  \BibitemOpen
  \bibfield  {author} {\bibinfo {author} {\bibfnamefont {S.}~\bibnamefont
  {Boyd}}, \bibinfo {author} {\bibfnamefont {S.~P.}\ \bibnamefont {Boyd}},\
  and\ \bibinfo {author} {\bibfnamefont {L.}~\bibnamefont {Vandenberghe}},\
  }\href@noop {} {\emph {\bibinfo {title} {Convex optimization}}}\ (\bibinfo
  {publisher} {Cambridge university press},\ \bibinfo {year}
  {2004})\BibitemShut {NoStop}%
\bibitem [{\citenamefont {Zhang}\ and\ \citenamefont
  {Schaeffer}(2019)}]{zhang2019convergence}%
  \BibitemOpen
  \bibfield  {author} {\bibinfo {author} {\bibfnamefont {L.}~\bibnamefont
  {Zhang}}\ and\ \bibinfo {author} {\bibfnamefont {H.}~\bibnamefont
  {Schaeffer}},\ }\bibfield  {title} {\bibinfo {title} {On the convergence of
  the sindy algorithm},\ }\href@noop {} {\bibfield  {journal} {\bibinfo
  {journal} {Multiscale Modeling \& Simulation}\ }\textbf {\bibinfo {volume}
  {17}},\ \bibinfo {pages} {948} (\bibinfo {year} {2019})}\BibitemShut
  {NoStop}%
\end{thebibliography}%

\end{document}